# Toward a Classification of Stochastically Stable Quenched Measures

Pierluigi Contucci

Dipartimento di Matematica
Università di Bologna,
I-40127, Bologna, Italy

contucci@dm.unibo.it

(May the 6th, 2002)

**Abstract**

In this short note we study the fourth order consequences of the stochastic stability property for mean field spin glass models introduced in [AC]. We show that due to a remarkable cancellation mechanism it reduces to the well known second order version as predicted by Parisi's *replica-equivalent* ansatz.

# 1 Introduction

In recent times a considerable amount of effort has been devoted to clarify the statistical mechanics structure of those models that describe disordered systems like the spin glass. A common feature of the considered class of models is the presence of two time scales, with different order of magnitude, governing the microscopic dynamics which reflects on a precise ensemble prescription in equilibrium statistical mechanics commonly called *quenched ensemble*.

Among the different approaches to the problem the so called *mean field* considers systems in which every particle interacts with the remaining ones with the same random strength (see [SK], [MPV], [TAL] and [B]).

In [AC] a stability property for the mean field models was investigated and its meaning in terms of continuity (in the temperature) of the thermodynamic functions was clarified. In particular some of the consequences of such a stability were proved to be captured by an infinite family of constraint for the overlap distributions. The result, in full agreement with the Parisi solution, provided an infinite family of overlap polynomials with zero average. Similar results were obtained in [G] and later generalized in [GG].

The stability property, has been then investigated in [FMPP1] and [FMPP2] and cleverly used to determine a relation between the off-equilibrium dynamics and the static properties. Among other things such a relation allows an experimental (and numerical) measurement of the equilibrium order parameter for spin glass models and answers positively, at least in the mean field case, the important paramount question of whether the equilibrium properties of a glassy system have any relation at all with the off-equilibrium properties, the only ones physically accessible.

It is then of crucial importance to have a clear understanding of the stability property. The study can be approached at least from two different points of view. One, in the framework of the GREM models, is the study of the random measures over fields of points on the line with the property of invariance under a reshuffling dynamics (see [R], [BS] and [AR]).

The second, that we approach in this paper, deals with the characterization of the stability property in terms of the overlaps distribution.

The paper is organized as follow: section 2 introduces the model and states the main result in proposition (2.3). Section 3 contains its proof.

## 2  Definitions and Results

The SK spin glass model [SK],[MPV] has spin variables $\sigma_i = \pm 1$, $i = 1, \ldots, N$, interacting via the Hamiltonian

$$H_N(J, \sigma) = -\frac{1}{\sqrt{N}} \sum_{1 \leq i < j \leq N} J_{i,j} \sigma_i \sigma_j , \qquad (2.1)$$

where the $J_{i,j}$ are $\frac{N(N-1)}{2}$ independent normal Gaussian variables.

Using the symbol $Av$ to denote the Gaussian integration it is possible to prove [GT] that the quenched free energy density defined by

$$-\beta f_N(\beta) = \frac{1}{N} Av(\log Z(\beta, J)) \qquad (2.2)$$

converges (increasing monotonically) in the $N \to \infty$ limit.

If we rescale an independent sample of the energy $H$ by a factor $\sqrt{N}$ we define the

$$K_N(J', \sigma) = -\frac{1}{N} \sum_{1 \leq i < j \leq N} J'_{i,j} \sigma_i \sigma_j , \qquad (2.3)$$

where $J'_{i,j}$ are i.i.d. unit Gaussian independent from $J_{i,j}$ of which turns out to be a Gaussian family indexed by $\sigma$ whose covariance is the positive definite matrix:

$$q^2(\sigma, \sigma') = \left[\frac{1}{N} \sum_i \sigma_i \sigma'_i\right]^2 . \qquad (2.4)$$

It is extremely useful to introduce the quenched measure over an arbitrary number of copies of the system. For such a purpose we may consider $r$ copies of the same system: each system is given the Hamiltonian $H^{(l)}$ where $l$ ranges from 1 to $r$. To each copy we associate the Random-Gibbs-Bolztmann state given by the usual formula:

$$< - >^{(l)} (-) = \frac{\sum_\sigma - e^{-\beta H^{(l)}(\sigma)}}{\sum_\sigma e^{-\beta H^{(l)}(\sigma)}} . \qquad (2.5)$$

$$<< - >> = \otimes_{l=1}^r < - >^{(l)} . \qquad (2.6)$$

The full measure $E(-)$ is then defined as

$$E(-) = Av(\otimes_{l=1}^r < - >^{(l)}) . \qquad (2.7)$$

The matrix $Q$, whose elements are defined by:

$$Q_{l,m} = q^2(\sigma^{(l)}, \sigma^{(m)}) \qquad (2.8)$$

is then a symmetric random matrix with respect to the measure $E(-)$ because its distribution turns out to be invariant under the action of the permutation group over the $r$ copies. For instance for $r = 3$

$$Q = \begin{pmatrix} 1 & Q_{1,2} & Q_{1,3} \\ Q_{2,1} & 1 & Q_{2,3} \\ Q_{3,1} & Q_{3,2} & 1 \end{pmatrix} \tag{2.9}$$

and

$$E(Q_{1,2}Q_{2,3}) = Av\{[Z(\beta,E)]^{-3} \sum_{\sigma^{(1)},\sigma^{(2)},\sigma^{(3)}} q^2(\sigma^{(1)},\sigma^{(2)}) q^2(\sigma^{(2)},\sigma^{(3)}) \cdot$$
$$\cdot e^{\beta(H(\sigma^{(2)},J)+H(\sigma^{(2)},J)+H(\sigma^{(3)},J))}\}. \tag{2.10}$$

We define as in [AC] the Random-Gibbs-Boltzmann deformed state as

$$< - >_{\lambda K} \quad := \quad \frac{< - \exp \lambda K >}{< \exp \lambda K >},$$

$$\ll - \gg_{\lambda K} \quad := \quad \otimes_l < - >_{\lambda K}^{(l)}, \tag{2.11}$$

and

$$E_{\lambda K}(-) := Av(\ll - \gg_{\lambda K}). \tag{2.12}$$

It may be shown that the Wick integration formula for Gaussian systems imply that all the *thermal* observables of the SK model (i.e the derivatives of the free energy with respect to $\beta$) are indeed moments of the Random matrix $Q$ with respect to its permutation invariant measure $E$.

In [AC] it was shown the stochastic stability property:

**Proposition 2.1** *If, over a certain temperature range, the states $E^{(N,\beta)}(-)$ are uniformly continuous in beta, as $N \to \infty$, then the limiting measures $E^{(\beta)}(-)$ are stable under deformations in $\lambda K$, in the sense:*

$$E(-) = E_{\lambda K}(-). \tag{2.13}$$

Always in [AC] it was derived as a consequence of the stochastic stability that an infinite family of overlap polynomial has zero mean. In fact it was proved that:

**Proposition 2.2** *At finite values of $N$ and for every overlap monomial $M$*

$$\frac{\partial}{\beta\partial\beta}E_N(M) = N\frac{\partial^2}{\partial\lambda^2}E_{N,\lambda K}(M)|_{\lambda=0} = NE_N(\Delta M), \tag{2.14}$$

*where $\Delta M$ is a suitable overlap polynomial. In particular, if the quenched state is locally differentiable in $\beta$ uniformly in $N$, the expectation values of quantities of the form $\Delta M$ vanish at the rate $O(1/N)$ or faster i.e. in the limit*

$$\frac{\partial^2}{\partial\lambda^2}E_{\lambda K}(M)|_{\lambda=0} = E(\Delta M) = 0 . \tag{2.15}$$

**Remark**: *the family of polynomials $\Delta M$ whose mean is zero plays in the mean field spin glass models a role which is reminiscent of the Hermite polynomials in classical Gaussian theories; in the same way the identities that can be found in [GG] allow a computation of the moments in terms of sum of products of lower order ones and are, in parallel, reminiscent of the Wick formula for the Gaussian case.*

In this short note we want to investigate if stochastic stability when considered for higher order derivatives in $\lambda$ lead to new results with respect to its second order consequences as one may a priori expect or if it is rather already captured by the second order as conjectured by Parisi in his replica equivalent ansatz (a generalization of the replica symmetry breaking ansatz [P]). For this purposes we consider here the fourth order derivative with respect to $\lambda$ (the *odd*-order derivatives are zero for parity, the general *even*-order case will be considered in [CK]) and we show the following results:

**Proposition 2.3** *For every overlap monomial $M$*

$$\frac{\partial^4}{\partial\lambda^4}E_{\lambda K}(M)|_{\lambda=0} = 3E(\Delta^2 M) . \tag{2.16}$$

As a corollary of (3.25) and (2.16) and considering the fact that $\Delta M$ is a polynomial in the overlap algebra i.e. $\Delta M = \sum_\alpha c_{\alpha,M} M_\alpha$ the vanishing of the average of $\Delta M$ for every $M$ implies also the vanishing of the average of $\Delta^2 M$:

$$\frac{\partial^4}{\partial\lambda^4}E_{\lambda K}(M)|_{\lambda=0} = 3E(\Delta^2 M) = 3\sum_\alpha c_{\alpha,\Delta M}E(\Delta M_\alpha) = 0 , \tag{2.17}$$

where the last equality holds due to the (2.15). In conclusion the consequences of stochastic stability at the fourth order are included in the consequences at the second order as predicted by the *replica-equivalent* Parisi ansatz [P]. An easy but important corollary of the previous proposition is the following

**Proposition 2.4** *In [AC] it was proved that*

$$\left(\frac{\partial}{\beta\partial\beta}\right)^2 E_N(M) = N^2 \frac{\partial^4}{\partial\lambda^4} E_{N,\lambda K}(M)|_{\lambda=0} , \qquad (2.18)$$

*due to the (2.17) we have*

$$\left(\frac{\partial}{\beta\partial\beta}\right)^2 E_N(M) = 3N^2 E_N(\Delta^2 M) . \qquad (2.19)$$

*If, over a certain temperature range, the states $E_N(-)$ are uniformly differentiable in beta $(C_1)$, as $N \to \infty$, then over the same range (up to, at most, a countable number of points) they are uniformly $(C_2)$ and the expectation values of quantities of the form $\Delta^2 M$ vanish at the rate $O(1/N^2)$ or faster.*

## 3  Proof of proposition 2.3

In the sequel we will have to deal with derivatives of functions involving Gaussian variables with covariance $Q$. The functions we will consider will be products of Gaussian and the Boltzmann weights. A typical quantity could be $Q_{1,2}^2 Q_{2,3}$. Computing the $\lambda$-derivatives of $E(Q_{1,2}^2 Q_{2,3})_\lambda$ will amount to consider repeated *truncated* expectations. It may be useful then to introduce a graphical representation in which a monomial of the previous form is identified with a graph whose vertexes are the replica indices $\{1, 2, 3, ...r\}$ and the edges correspond to the overlaps $Q_{l,m}$. Such a graph will be indicated by the symbol $(1,2)^2(2,3)$. The treatment of *unpaired* Gaussian variables will be denoted by *free legs* and denoted by symbols like $(1)$, $(2)$ etc. It is important to stress that for each monomial in the overlap variables the invariance under permutation of $(E, Q)$ will reflect on invariance under permutation labeling over graphs: $(1,2)(2,3) = (2,1)(1,3) = (2,15)(15,3)$. Moreover we will only consider the $(l,l) = 1$ case, which covers all the $p$ spin models, and Derrida's REM and GREM structures. On a mathematical point of view it is then convenient the introduction of the notion of *generalized* graphs which is an object containing both *edges* and *legs*: $(1,2)^2(3)$, $(1,2)(3,4)^2(1)$. The permutation symmetry over indices will imply for instance $(1,2)^2(3) = (5,6)^2(1)$, $(1,2)(2,3)(3,1)(1)^2 = (3,4)(3,2)(2,4)(3)$ etc.

We introduce the linear operation $\delta$ which acts on generalized graphs as:

$$\delta G = \sum_{v \in V(G)} \delta_v G , \qquad (3.20)$$

$$\delta_v G = \delta_v^{(+)} + \delta_v^{(-)} , \qquad (3.21)$$

$$\delta_v^{(+)} G = (v)G, \qquad (3.22)$$

**Ex:** $\delta_2^{(+)}(1,2) = (2)(1,2)$.

$$\delta_v^{(-)} G = -(v')G \qquad (3.23)$$

where $v'$ is any index not belonging to $G$.

**Ex:** $\delta_3^{(-)}(1,3) = -(2)(1,3)$, $\delta_3^{(-)}\delta_3^{(-)}(1,3) = (4)(2)(1,3)$,

We introduce then the *Wick* contraction map $\mathcal{C}$ which acts only on the *legs* part of the graph according to the Wick formula for Gaussian

Let $I$ be set of indices with *even* cardinality, then

$$\mathcal{C} \prod_{i \in I}(i) = \sum_{pairings} \prod_{v,v' \in I}(v,v'). \qquad (3.24)$$

Ex: $\mathcal{C}[(1)(2)(3)(4)] = 3(1,2)$

**Proposition 3.5** *For every generalized graph $G$*

$$\frac{\partial^4}{\partial \lambda^4} E_{\lambda h}(G)|_{\lambda=0} = E(\mathcal{C}\delta^4 G), \qquad (3.25)$$

*where $\mathcal{C}\delta^4 G$ is a polynomial in the matrix entries.*

The proof of the above proposition is straightforward. The operation $\delta$ is the graphical counterpart of the usual derivative with respect to the parameter $\lambda$ in the Boltzmann weight (where it appears in $\lambda K$). Such a derivative produces a *truncated correlation* expressed in the rule (3.21). The (3.20) is nothing but the Leibniz rule for derivative of products. Each differentiation with respect to $\lambda$ produces a sum of monomials, each containing an added centered Gaussian variable ($K$) of zero mean. The contraction is nothing but the Wick rule for normalized Gaussian families.

We will get the result (3.25) as a direct consequence of the following combinatorial theorem:

**Proposition 3.6** *Defining $\Delta := \mathcal{C}\delta^2$,*

$$\mathcal{C}\delta^4 = 3\Delta^2. \qquad (3.26)$$

**Remark:** *it is interesting to note that the left hand side of the last equation contains a priori a number of addends which is $3 \cdot 16 = 48$ (the 3 coming from the Wick contraction of a fourth order monomial and the 16 coming from $2^4$ terms of $(\delta^+ + \delta^-)^4$ ). The* *right hand side instead contains* $4 \cdot 4 = 16$ *terms from the square of* $\Delta$. *Although the Wick contraction doesn't conserve the number of edges nor the number of vertexes (see formulas 3.44 to 3.47 in the next section) the presence of alternating signs in the definition of* $\delta$ *together with the invariance of permutation of graphs labeling will produce a delicate cancellation mechanism responsible for the clean equality (3.6).*

**Proof.** Let show, for the sake of simplicity, the theorem for the moment $E(Q_{1,2})$. The generalization will be discussed at the end.

As in [AC] we have

$$\frac{1}{2}\Delta(1,2) = (1,2)^2 - 4(1,2)(2,3) + 3(1,2)(3,4) \tag{3.27}$$

Applying the $\Delta$ to the previous formula addends we obtain

$$\frac{1}{2}\Delta[(1,2)^2] = (1,2)^3 - 4(1,2)^2(1,3) + 6(1,2)(3,4), \tag{3.28}$$

$$\frac{1}{2}\Delta[(1,2)(2,3)] = \; + \; 2(1,2)^2(2,3) - 9(1,2)(2,3)(3,4) + (1,2)(2,3)(3,1)$$
$$- \; 3(1,2)(1,3)(1,4) + 6(1,2)(1,3)(4,5) \tag{3.29}$$

$$\frac{1}{2}\Delta[(1,2)(3,4)] = \; + \; 2(1,2)^2(3,4) + 4(1,2)(2,3)(3,4) - 16(1,2)(1,3)(4,5)$$
$$+ \; 10(1,2)(3,4)(5,6). \tag{3.30}$$

Summing the three contributions we have

$$\frac{1}{4}\Delta^2(1,2) = \; + \; 2(1,2)^3 - 24(1,2)^2(1,3) - 8(1,2)(2,3)(3,1)$$
$$+ \; 18(1,2)^2(3,4) - 144(1,2)(1,3)(4,5) + 96(1,2)(2,3)(3,4)$$
$$+ \; 60(1,2)(3,4)(5,6) + 24(1,2)(1,3)(1,4). \tag{3.31}$$

The left hand side of (3.6) is:

$$\frac{1}{2}\delta^2[(1,2)] = (1,2)(1)(2) - 4(1,2)(1)(3) + (1,2)(3)(4) + (1,2)(1)^2 - (1,2)(3)^2. \tag{3.32}$$

Applying the $\delta^2$ to the each term we get:

$$\frac{1}{2}\delta^2[(1,2)(1)(2)] = \; + \; (1,2)(2)^3(1) + (1,2)(2)^2(1)^2 - 4(1,2)(1)(2)^2(3)$$
$$- \; (1,2)(1)(2)(3)^2 + 3(1,2)(1)(2)(3)(4) \tag{3.33}$$

$$\frac{1}{2}\delta^2[(1,2)(1)(3)] = \quad + \quad 3(1,2)(1)^2(2)(3) + 2(1,2)(1)(2)(3)^2 - 6(1,2)(1)(2)(3)(4)$$
$$+ \quad (1,2)(1)^3(3) + 2(1,2)(1)^2(3)^2 - 6(1,2)(1)^2(3)(4)$$
$$- \quad 9(1,2)(1)(2)^2(3) + (1,2)(1)(3)(4)(5) \tag{3.34}$$

$$\frac{1}{2}\delta^2[(1,2)(3)(4)] = \quad + \quad 2(1,2)(1)^2(3)(4) + 2(1,2)(1)(2)(3)(4) + 8(1,2)(1)(3)^2(4)$$
$$- \quad 16(1,2)(1)(3)(4)(5) + 2(1,2)(3)^3(4) + 2(1,2)(3)^2(4)^2$$
$$- \quad 20(1,2)(3)^2(4)(5) + 20(1,2)(3)(4)(5)(6) \tag{3.35}$$

$$\frac{1}{2}\delta^2[(1,2)(1)^2] = \quad + \quad (1,2)(1)^2(2)^2 + 2(1,2)(1)(2)^3 - 4(1,2)(1)(2)^2(3) + (1,2)(1)^4$$
$$- \quad 4(1,2)(1)^3(3) - 2(1,2)(2)^2(3)^2 + 6(1,2)(1)^2(3)(4) \tag{3.36}$$

$$\frac{1}{2}\delta^2[(1,2)(3)^2] = \quad + \quad 2(1,2)(2)^2(3)^2 + 2(1,2)(1)(2)(3)^2 + 4(1,2)(1)(3)^3$$
$$- \quad 12(1,2)(1)(3)^2(4) + (1,2)(3)^4 - 6(1,2)(3)^3(4)$$
$$- \quad 3(1,2)(3)^2(4)^2 + 12(1,2)(3)^2(4)(5) \,. \tag{3.37}$$

We remind that at the fourth order the Wick rule produces 4 different topological graphs. Considering the invariance under permutation of the graphs labels we obtain:

$$\mathcal{C}[(1)(2)(3)(4)] = 3(1,2)(3,4) \,, \tag{3.38}$$
$$\mathcal{C}[(1)^2(2)(3)] = 2(1,2)(2,3) + (1,2) \,, \tag{3.39}$$
$$\mathcal{C}[(1)^2(2)^2] = 2(1,2)^2 + 1 \,, \tag{3.40}$$
$$\mathcal{C}[(1)^3(2)] = 3(1,2) \,. \tag{3.41}$$

We may now sum up the 5 contributions and apply the wick contraction $\mathcal{C}$ to the result. We obtain:

$$\frac{1}{4}\mathcal{C}[\delta^4(1,2)] = 3 \quad [\quad 2(1,2)^3 - 24(1,2)^2(1,3) - 8(1,2)(2,3)(3,1) + 18(1,2)^2(3,4)$$
$$- \quad 144(1,2)(1,3)(4,5) + 96(1,2)(2,3)(3,4) + 60(1,2)(3,4)(5,6)$$
$$+ \quad 24(1,2)(1,3)(1,4)]$$
$$= \quad 3\frac{1}{4}[\Delta^2(1,2)] \,, \tag{3.42}$$

which proves the theorem (3.6) when the two operators in $\mathcal{C}\delta^4$ and $\Delta^2$ are applied to the graph $(1,2)$. The extension to the general case is straightforward and already included in the well definiteness of the operators $\delta$ and $\Delta$ for general graphs as it was shown in section 6 of [AC]. The (3.6) immediately imply the (2.16).

**Acknowledgments.** We acknowledge useful conversations with Francesco Guerra, Andreas Knauf and Paola Trevisan.